------------------------------------------------------------------------------

Appended: paper in LaTeX format (one file) and postscript figures (one file)
          (see 'cut here')

--- Cut here ------------------------------------------------------------------

\documentstyle[12pt]{article}

\newcommand{\newsection}{ \setcounter{equation}{0} \section}
\newcommand{\beq}{\begin{equation}}
\newcommand{\eeq}{\end{equation}}
\newcommand{\bea}{\begin{eqnarray}}
\newcommand{\eea}{\end{eqnarray}}

\newcommand{\be}{\begin{enumerate}}
\newcommand{\ee}{\end{enumerate}}
\newcommand{\bi}{\begin{itemize}}
\newcommand{\ei}{\end{itemize}}
\newcommand{\ba}{\begin{array}}
\newcommand{\ea}{\end{array}}
\newcommand{\bc}{\begin{center}}
\newcommand{\ec}{\end{center}}
\newcommand{\bt}{\begin{tabular}}
\newcommand{\et}{\end{tabular}}
\newcommand{\half}{\textstyle {1\over2} \displaystyle}    
\newcommand{\third}{\textstyle {1\over3} \displaystyle}   
\newcommand{\quarter}{\textstyle {1\over4} \displaystyle} 
\newcommand{\sixth}{\textstyle {1\over6} \displaystyle}   
\newcommand{\tenth}{\textstyle {1\over10} \displaystyle}  
\newcommand{\twoth}{\textstyle {2\over3} \displaystyle}   
\newcommand{\Dslash}{{\hbox{D}\kern-0.6em\raise0.15ex\hbox{/}}} 

\renewcommand{\d}{\delta}

\newcommand{\cR}{{\cal R}}

\newcommand{\noi}{\noindent}

\begin{document}
\topmargin 0pt
\oddsidemargin 5mm
\headheight 0pt
\topskip 0mm

\addtolength{\baselineskip}{0.20\baselineskip}
\hfill DAMTP-93-27

\hfill UCI-Th-93-16

\hfill June 1993

\begin{center}

\vspace{36pt}
{\Large \bf Simplicial Gravity Coupled to Scalar Matter }

\vspace{24pt}

{\sl Herbert W. Hamber
\footnote{Supported in part by the National Science
Foundation under grant PHY-9208386}  }

\vspace{12pt}

Department of Physics \\
University of California at Irvine \\
Irvine, California 92717 USA \\

\vspace{18pt}

{\sl Ruth M. Williams}

\vspace{12pt}

Department of Applied Mathematics and Theoretical Physics \\
Silver Street \\
Cambridge CB3 9EW, England \\

\end{center}

\vfill

\begin{center} {\bf ABSTRACT } \end{center}
\vspace{12pt}
\noi

A model for quantized gravity coupled to matter in the form of a single
scalar field is investigated in four dimensions.
For the metric degrees of freedom we
employ Regge's simplicial discretization, with the scalar fields
defined at the vertices of the four-simplices. We examine how the
continuous phase transition found earlier,
separating the smooth from the rough phase of quantized gravity,
is influenced by the presence of scalar matter.
A determination of the critical exponents seems to indicate that the
effects of matter are rather small, unless the number of scalar flavors is
large. Close to the critical point where the average curvature
approaches zero, the coupling of matter to gravity is found to be weak.
The nature of the phase diagram and the values for the critical
exponents suggest that gravitational interactions increase with
distance.

\vspace{24pt}
\vfill
\newpage

\vskip 10pt
\newsection{Introduction}

Any serious attempt at understanding the ground state properties
of quantized gravity has to include at some stage the consideration
of the effects of matter fields. While there are many choices
for the matter fields and for their interactions, the simplest
actions to deal with in the framework of a lattice model for gravity
are the ones that represent one (or more) scalar fields.
In this paper we will discuss a first attempt at determining those
effects.

Regge's model is the natural discretization for quantized gravity
\cite{regge}.
At the classical level, it is completely equivalent to general relativity,
and the correspondence is particularly transparent in the lattice
weak field expansion, with the invariant edge lengths playing
the role of infinitesimal geodesics in the continuum.
In the limit of smooth manifolds with small curvatures, the
continuous diffeomorphism invariance of the continuum theory
is recovered \cite{rw,hw3d}.
But in contrast to ordinary lattice gauge theories, the model is formulated
entirely in terms of coordinate invariant quantities, the edge
lengths, which form the elementary degrees of freedom in the
theory \cite{hartle,rw1}.

Recent work based on Regge's simplicial formulation of gravity has shown,
in pure gravity without matter,
the appearance in four dimensions of a phase transition in the
bare Newton's constant, separating a smooth phase with small negative
average curvature from a rough phase with large positive
curvature \cite{tala,ph1}.
While the fractal dimension is rather small in the rough phase,
indicating a tree-like geometry for the ground state, it is very close
to four in the smooth phase close to the critical point.
Furthermore, a calculation of the critical exponents in the smooth
phase close to the critical point indicates that the transition is apparently
second order with divergent curvature fluctuations, and that a
lattice continuum can be constructed.

Very similar results have recently been obtained in the dynamical
triangulation model for gravity, in the sense that a similar phase
transition was found separating what appear to be the same type
of phases \cite{dt4d}.
This development represents an alternative and complementary
approach to what is being discussed here.
However it has not been possible yet in these models to extract the
critical exponents, and it is therefore not clear
yet whether a continuum limit really exists. In particular it appears
that close to the transition, the dynamical triangulation model
does not give rise to the correct scaling properties for the curvature,
which are necessary to define a lattice continuum limit.
It is therefore unclear whether the transition is
first order as a consequence of the discreteness of the
curvatures, with no continuum limit (as one finds for
example in lattice gauge theories based on discrete subgroups of
$SU(N)$ \cite{br}).
While in two dimensions both lattice models lead to similar results
both in the absence and presence of scalar matter \cite{dt2d,hw2d,gh},
in three dimensions the
dynamical triangulation model has no continuum limit \cite{dt3d},
in apparent disagreement with the continuum expectations \cite{wei,kn},
and the simplicial Regge gravity results \cite{hw3d}, which suggest instead
that a well defined continuum limit exists (albeit trivial in
the absence of matter, with the scalar curvature playing the role of a
scalar field). These results are rather disappointing, since
it would be desirable to have two rather different, independent
discretizations for gravity, with the same lattice continuum limit.
It is not clear yet at this point whether these results indicate
a fundamental flaw in the model (lack of restoration of broken diffeomorphism
invariance), or simply a perhaps surmountable technical difficulty in
determining exponents. For a clear recent review of some of these
aspects in the dynamically triangulated models we refer the reader
to the last reference in \cite{dt4d}.

In this paper we will present some first result on the properties of
Regge's simplicial gravity coupled to a scalar field, as derived from
numerical studies on lattices of up to $24 \times 16^4 = 1,572,864$
simplices.
The paper is organized as follows. First we
discuss in Sec. 2 the simplicial action and measure for the combined
gravitational and scalar degrees of freedom. Then we digress in Sec. 3
on what is known about the effects of scalar matter fields in the continuum,
to the extent that the results will be relevant for our later
calculations. We then present in Sec. 4 the definition of physical
observables which can be measured when scalar fields are present,
besides the purely gravitational ones introduced previously, and how
these can be related to effective low energy couplings.
In Sec. 5 we present our results and their interpretation, and in
Sec. 6 we give a discussion on how other quantities such as the curvature
and volume distributions can be obtained close to the critical point.
Sec. 7 then contains our conclusions.

\vskip 10pt
\newsection{Action and Measure for the Scalar Field}

Following \cite{hw84}, the four-dimensional pure gravity action on the
lattice is written as
\beq
I_g [l] =  \sum_ {\rm hinges \, h } \Bigl [ \, \lambda \, V_h -
k \, A_h \d _h + a \, { A_ h^2  \delta _ h^2 \over  V _ h } \, \Bigr ] ,
\label{eq:acg}
\eeq
where $V_h$ is the volume per hinge (represented by a triangle in four
dimensions),
$A_h$ is the area of the hinge and $\delta_h$ the corresponding
deficit angle, proportional to the curvature at $h$.
The term proportional to $k$ is the original Regge action. In the lattice
weak field expansion, the last two terms both contain higher derivative
contributions \cite{rw,hw3d} (in the last term it is the leading
contribution).
This is a simple consequence of the fact that on the lattice finite
differences give rise, when Fourier transformed, to terms involving
trigonometric functions of the lattice momenta.
The higher order corrections are
in general expected to be irrelevant in the continuum limit, if one
can be found, and unless the coefficient $a$ is taken to be very large
in this limit.
Whenever systematic studies have been done, there are
indications that this is indeed the case \cite{gh,hw3d}, as one would
expect from the experience gained in other, simpler model field theories.
The results of ref. ~\cite{ph1} in four dimensions
also suggest that the corrections are
negligible in the lattice continuum limit ($k \rightarrow k_c $),
and that the `ghost mass' associated with the higher derivative
corrections remains of the order of the ultraviolet cutoff,
of the order of the inverse average lattice spacing,
$m_{ghost} \sim \pi / l_0 $  (for a general
discussion of some of these points in simpler field theory models, see
for example \cite{pbook}). In the context of the present work the
higher derivative terms will be considered as convenient invariant
regulators, in addition to the usual lattice cutoff.

In the classical continuum limit the above action is
equivalent \cite{rw,hw3d,lee,cms,hw84} to
\beq
I_g [g] = \int  d^4 x  \, \sqrt g \, \Bigl [ \, \lambda - \half k \, R
+ \, \quarter a \, R _ { \mu \nu \rho \sigma }  R ^ { \mu \nu \rho \sigma }
+ \cdots \, \Bigr ] ,
\label{eq:acgc}
\eeq
with a cosmological constant term (proportional to $\lambda$), the
Einstein-Hilbert term ($k = 1 / ( 8 \pi G ) $), and a higher derivative
term, and with the dots indicating higher order lattice corrections.
In the following we will follow the convention of choosing the fundamental
lattice spacing
to be equal to one; the correct power of the lattice spacing needed
to convert lattice to continuum quantities can always be restored
by invoking dimensional arguments (but we have to remember that
due to the dynamical nature of the lattice, the average distance between
sites, in units of the fundamental lattice spacing, will still depend on
the bare couplings and the measure).
For an appropriate choice of bare couplings,
the above lattice action is bounded below for a regular lattice, even
for $a=0$, due to the presence of the lattice momentum cutoff \cite{rw}.
For non-singular measures and in the presence of the $\lambda$-term
such a regular lattice can be shown to arise naturally.
The higher derivative terms can be set to zero ($a=0$),
but they nevertheless seem to be necessary for reaching the lattice continuum
limit, and are in any case generated by radiative corrections
already in weak coupling perturbation theory.
When scalar fields are introduced, higher derivative terms are
generated as well by the quantum fluctuations of the scalar field.
Renormalization group arguments then suggest that in general the continuum
limit should be explored in this enlarged multi-parameter space.
Some very interesting suggestions regarding properties of non-renormalizable
theories beyond perturbation theory have been put forward in \cite{parisi}.

Next a scalar field is introduced, as the simplest type of dynamical
matter that can be coupled to gravity.
Consider an $n_f$-component field $\phi_i^a$, $a=1,...,n_f$, and
define this field at the vertices of the simplices.
Introduce finite lattice differences defined in the usual way
\beq
( \Delta _ \mu  \phi^a )_i =
{ \phi^a _{ i + \mu } - \phi^a _ i \over l _ { i,i+ \mu } } .
\eeq
The index $\mu$ labels the possible directions in which one can move from
a point in a given triangle, and
$l _ { i , i + \mu }$ is the length of the edge connecting the two points.
For simplicity let us consider for now the case $n_f=1$.
Then add to the above discrete pure gravitational action the contribution
\beq
I_\phi [l, \phi ] = \half \sum_{<ij>} V_{ij} \,
\Bigl ( { \phi_i - \phi_j \over l_{ij} } \Bigr )^2 \, +
\half \sum_{i} V_{i} \, (m^2 + \xi R_i ) \phi_i^2  +
\sum_{i} V_{i} \, U( \phi_i )  + \cdots ,
\label{eq:acp}
\eeq
where $U(\phi)$ is a potential for the scalar field, and the term containing
the discrete analog of the scalar curvature involves
\beq
V_{i} R_i \equiv \sum_{ h \supset i } \delta_h A_h  \sim \sqrt{g} R
\eeq
In the expression for the scalar action,
$V_{ij}$ is the volume associated with the edge $l_{ij}$,
while $V_i$ is associated with the site $i$.
There is more than one way to define such a volume \cite{hw84,cfl,bi}, but
under reasonable assumptions, such as positivity,
one should get equivalent results in the continuum.
The agreement between different lattice actions in the smooth limit
can be shown explicitly in the lattice weak field expansion, but the
calculations can be rather tedious and we will present the results elsewhere.
Here we will restrict ourselves to the baricentric volume
subdivision \cite{hw84} which is the simplest to deal with.
The above lattice action then corresponds to the continuum expression
\beq
I_\phi [ g, \phi ] = \half \int \, \sqrt g \; [  \,
g ^ { \mu \nu } \, \partial _ \mu  \phi \, \partial _ \nu  \phi
+ ( m^2 + \xi  R ) \phi^2   ] + \int \, \sqrt g \,  U( \phi ) + \cdots ,
\eeq
with the induced metric related in the usual way to the edge lengths
\cite{rw,hw3d}.
As is already the case for the purely gravitational action,
the correspondence
between lattice and continuum operators is true classically only up to higher
derivative corrections. But such higher derivative corrections in the scalar
field action are expected to be irrelevant and we will not consider them
here any further.
The scalar field potential $U(\phi)$ could contain quartic contributions,
whose effects are of interest in the context of cosmological
models where spontaneously broken symmetries play an important role.
For the moment we will be considering a scalar field
without direct self-interactions, and will set $U$=0.

The lattice scalar action contains a mass parameter $m$, which has
to be tuned to zero in lattice units to achieve the lattice continuum
limit for scalar correlations.
The dimensionless coupling $\xi$ is arbitrary;
two special cases are the minimal ($\xi = 0$) and the conformal
($\xi = \sixth $) coupling case.
As an extreme case one could consider a situation in which
the matter action by itself is the only action contribution, without
any kinetic term for the gravitational field, but still with a
non-trivial
gravitational measure; integration over the scalar field would
then give rise to an effective non-local gravitational action.

Having discussed the action, let us turn now to the measure.
The discretized partition function can be written as
\beq
Z = \int d \mu [ l ] \, d \mu [ \phi ] \,
\exp \, \left \{ -I_g [ l ] -I_\phi [ l, \phi ] \right \} .
\eeq
It is well known that the continuum gravitational measure
is not unique, and different regularizations will lead to different
forms for the measure.
DeWitt has argued that the gravitational measure should have the
form \cite{dw,dwm}
\beq
\int d \mu [g] \, = \, \int \prod_x  g^{ (d-4)(d+1)/8 }
\prod_{\mu \ge \nu} d g_{\mu \nu} .
\eeq
The main difference between various Euclidean
measures seems to be in the power of
$\sqrt{g}$ in the prefactor, which on the lattice corresponds to some product
of volume factors.  On the lattice these volume factors do not give rise
to coupling terms, and are therefore strictly {\it local}.
It should also be clear that since diffeomorphism invariance is lost in
{\it all} lattice models of gravity, at least away from
smooth manifolds (the very definition of a lattice breaks
local Poincar\'e invariance), there is no clear criterion at this point
to help one decide which measure should be singled out.
We have argued before that the power appearing in the measure should
be considered as an additional, hopefully irrelevant, bare
parameter \cite{hw84}.

On the simplicial lattice the invariant edge lengths represent the
elementary degrees of
freedom, which uniquely specify the geometry for a given incidence matrix.
Since the induced metric at a simplex is linearly related to the edge
lengths squared within that simplex, one would expect the lattice
analog of the DeWitt metric to simply correspond to $dl^2$ \cite{hartle}.
We will therefore write the lattice measure as \cite{lesh,hw84,tala}
\beq
\int d \mu _ \epsilon [ l ] =
\prod _ {\rm edges \, ij}  \int_ 0 ^ \infty \,
V_{ij}^{ 2 \sigma} \, { d  l _ {ij} ^ 2 } \, F _ \epsilon [ l ] ,
\label{eq:meas}
\eeq
where $V_{ij}$ is the 'volume per edge',
$ F _ \epsilon [l] $  is
a function of the edge lengths which enforces the higher-dimensional
analogs of the triangle inequalities,
and $\sigma = 0$ for the lattice analog of the DeWitt measure.
The parameter $ \epsilon $ is introduced as an ultraviolet
cutoff at small edge lengths: the function $F _ \epsilon [l]$ is
zero if any of the edges are equal to or less than $\epsilon$.
In general it is needed for sufficiently singular measures; for the $\sigma=0$
measure such a parameter is not necessary since
the triangle inequalities already strongly suppress small edge
lengths \cite{ph1}, and so we will set it to zero.
Note therefore that {\it no} cutoff is imposed on small or large edge
lengths, if a non-singular measure such as $dl^2$ is used.
This fact is essential for the recovery of diffeomorphism invariance
close to the critical point, where on large lattices
a few rather long edges, as well as some rather short ones,
start to appear \cite{tala,ph1}.

Eventually it is of interest to
systematically explore the sensitivity of the results
to the type of gravitational measure employed.
This has been done to a certain extent
in two \cite{gh} and three \cite{hw3d} dimensions.
The conclusion seems to be that for non-singular measures
the results relevant for the lattice continuum limit
(i.e. the long distance properties of the theory, as characterized
for example by the critical exponents)
appear to be independent of $\sigma$.
{}From a general point of view it is difficult to see how local volume
factors, which involve no gradient terms,
can possibly affect the nature of the continuum limit, which
is expected to be dominated by shear-wave-like distortions of the geometry of
space-time. The experience gained so far
seems to indicate that the volume factors
coming from the measure will only affect the overall lattice scale and the
shape of the distribution for the edge lengths, and will lead therefore
to different renormalizations of the cosmological constant, but will leave
the long-wavelength excitation spectrum, which is determined by the
relatively small fluctuations in the edge lengths about the lattice
equilibrium position, unaffected. But of course these arguments
cannot be taken as a substitute for a systematic investigation of this
issue in four dimensions.

In the presence of matter, similar considerations apply.
If an $n_f$-component scalar field is coupled to gravity
the power $\sigma$ appearing in the measure has to be changed, due
to an additional factor of $\prod_{x} ( \sqrt{g} )^{n_f/2}$
in the continuum gravitational measure.
On the lattice one then has $ \sigma  = n_f /30 $,
since with our discretization of spacetime based on hypercubes
there are $2^d -1=15$ edges emanating from each lattice vertex.
The additional measure factor insures that
\beq
\int \prod_x \left \{ d \phi  ( \sqrt{g} )^{n_f / 2} \right \} \,
\exp \left ( - \half m^2 \int \sqrt{g} \, \phi^2 \right ) \, = \,
\left [ \left ( \frac{ 2 \pi}{m^2} \right )^{n_f/2} \right ]^V = const
\eeq
or that for large mass, the scalar field completely decouples, leaving
only the dynamics of the pure gravitational field.

\vskip 10pt
\newsection{Effects of Matter Fields}

As long as the scalar action is quadratic, one can formally integrate
out the matter fields and obtain an effective Lagrangean contribution
written entirely in terms of the metric field,
\bea
\int d \mu [ \phi ] \, e^{ - \half \int \, \sqrt g \, \phi M[g] \phi }
& \equiv & \int \prod_x \left \{ d \phi \, ( \sqrt{g} )^{n_f / 2} \right \}
\, \exp \left \{ - \half \int \, \sqrt g \, \phi M[g] \phi \right \}
\nonumber \\
& \sim & \left \{ \det M[g] \right \}^{-n_f / 2}
\sim e^{- I_{eff} [g] } .
\eea
Here we have from the scalar field action
\beq
< x \vert \, M[g] \, \vert y > \, \equiv \,
( - \partial^2 + \xi R + m^2 ) \, \delta (x-y) ,
\eeq
where $\partial^2$ is the usual covariant Laplacian,
\beq
\partial^2 \phi \equiv  \frac{1}{\sqrt{g}} \, \partial_\mu \sqrt{g} \,
g ^ { \mu \nu } \partial_\nu  \phi .
\eeq
The full effective action, with terms from
Eq. ~(\ref{eq:acgc}) included, can be obtained from the results
of Ref. \cite{gi} (after introducing a proper time short distance
cutoff of the order of $s_0 \sim 1 / \Lambda^2 $). One finds then
\beq
I_{eff} [g] = \int \sqrt{g} \, \left [ \, \lambda' - \half k' \, R + \quarter
a' \, R_{\mu\nu\rho\sigma} R^{\mu\nu\rho\sigma} + \cdots \, \right ] ,
\eeq
with effective couplings (for one flavor, $n_f=1$)
\bea
\lambda' & = & \lambda + { 1 \over 64 \pi^2 } \, \Lambda^4
- { 1 \over 32 \pi^2 } \, m^2 \Lambda^2
+ { 1 \over 64 \pi^2 } \, m^4 \ln \Lambda^2 + \cdots
\nonumber \\
k' & = & k + { 1 \over 16 \pi^2 } (\xi - \sixth ) \, \Lambda^2
+ { 1 \over 16 \pi^2 } ( \xi - \sixth ) \, m^2 \ln \Lambda^2 + \cdots
\nonumber \\
a' & = & a + { 1 \over 1920 \pi^2 } \, \ln \Lambda^2 + \cdots .
\label{eq:effcoupl}
\eea
For a fixed cutoff these corrections
are quite small in magnitude compared to the corresponding gravitational
radiative corrections computed in the $2+\epsilon$ expansion \cite{wei,kn}
or in higher derivative theories \cite{hdqg}.
We will see later that this is also clearly the case for the lattice results.
As in ordinary gauge theories, matter vacuum polarization effects
are small unless one has a large number of matter fields
(in which case even a new phase might appear).
To the extent that the lattice scalar action is equivalent in the
lattice continuum limit to the corresponding continuum scalar action, the above
perturbative results, valid for small curvatures,
should be relevant for the lattice model as well.

The effects of matter fields are small also from the point of view
of the $2 + \epsilon$ perturbative expansion for gravity \cite{wei,kn}.
One analytically continues in the spacetime dimension by using
dimensional regularization, and applies perturbation theory about
$d=2$, where Newton's constant is dimensionless
(it is not clear whether this approach makes any sense beyond
perturbation theory).
In this expansion the dimensionful bare coupling is written as
$G_0 = \Lambda^{2-d} G $, where $\Lambda$
is an ultraviolet cutoff (corresponding on the lattice to a momentum
cutoff of the order of the inverse average
lattice spacing, $\Lambda \sim \, $$ \pi / <~l^2~>^{1/2}$)
and $G$ a dimensionless bare coupling constant.
A double expansion in $G$ and $\epsilon$
then leads in lowest order to
a nontrivial fixed point in $G$ above two dimensions,
where some local averages and their fluctuations
are expected to develop an algebraic singularity in $G$
(the problem of the unboundedness of the Euclidean gravitational
action does not appear in perturbation theory).
Close to two dimensions the gravitational
beta function is given to one loop by
\beq
\beta (G) \, \equiv \, { \partial G \over \partial log \Lambda } \, = \,
\epsilon \, G \, - \, \twoth (25- n_f ) \, G^2 \, + \cdots ,
\label{eq:beta}
\eeq
where $n_f$ is the number of massless scalar fields.
To lowest order the ultraviolet fixed point is at
\beq
G^* \, = \,
{ 3 \epsilon \over 2 (25 - n_f ) } \, + \, O( \epsilon^2 ) .
\eeq
Integrating Eq.~(\ref{eq:beta}) close to the non-trivial fixed point
in $2 + \epsilon $ dimensions we obtain
\beq
\mu_0 \, = \, \Lambda \, \exp \left ( { - \int^G \, {d G' \over \beta (G') } }
\right )
\, \mathrel{\mathop\sim_{G \rightarrow G^* }} \,
\Lambda \, | \, G - G^* |^{ - 1 / \beta ' (G^*) } \, \sim \,
\Lambda \, | \, G - G^* |^{ 1 / \epsilon } ,
\eeq
where $\mu_0$ is an arbitrary integration constant, with dimension
of a mass, and which should be identified with some physical scale.
The derivative of the beta function at the fixed point defines
the critical exponent $\nu$, which to this order is independent of $n_f$,
\beq
\beta ' (G^*) \, = \, - \epsilon \, = \, - 1/ \nu .
\eeq
The possibility of algebraic singularities
in the neighborhood of the fixed point, appearing in vacuum
expectation values such as the average curvature and its derivatives,
is then a natural one, at least from the point
of view of the $2+\epsilon$ expansion.

The previous results also illustrate
how in principle the lattice continuum limit should be taken \cite{pbook}.
It corresponds to $\Lambda \rightarrow \infty$,
$G \rightarrow G^*$ with $\mu_0$ held constant; for fixed lattice
cutoff the continuum limit is approached by tuning $G$ to $G^*$.
Alternatively, we can choose to compute dimensionless ratios
directly, and determine their limiting value as we approach the critical
point (we will show examples of this later).
Away from $G^*$ one will in general expect to encounter some
lattice artifacts, which reflect
the non-uniqueness of the lattice transcription of the continuum action
and measure, as well as its reduced symmetry properties.

Let us conclude this section by mentioning that the
Nielsen-Hughes formula \cite{nh} for the one-loop beta function associated
with a spin-$s$ particle in four dimensions provides for a physical
interpretation of the fact that the matter contribution is so small
compared to the gravitational one.
It appears that this result is related to the fact that the
spin of the graviton is not a small number.
Considering only spin 0 and 2, the formula gives the lowest order
result for the beta function coefficient as
\beq
16 \pi^2 \beta_0 = - \sum_s
(-1)^{2s} \, \left [ \, (2s)^2 - \third \, \right ]
= - \frac{1}{3 } \, \bigl ( 47 - n_f \bigr ) ,
\label{eq:nh}
\eeq
making the matter contribution quite negligible unless the number
of flavors is large.
In higher derivative theories one finds similar large coefficients.
It is encouraging that similar results are found
from the lattice calculations to be described below.
Furthermore, for a sufficiently
large number of flavors one would expect eventually a phase transition
(if these lowest order results are taken seriously), due
to the change of sign in the beta function.

\vskip 10pt
\newsection{ Observables }

When we consider gravity coupled to a scalar field, we can distinguish
two types of observables, those involving the metric field (the edge
lengths) only, and those involving also the scalar field.
Quantities such as the expectation value of the scalar curvature, the
fluctuations in the curvatures or
the curvature correlations belong to the first class, while scalar
field averages and scalar correlations belong to the second class.

Following \cite{tala}, we define the following gravitational physical
observables, such as the average curvature
\beq
\cR (\lambda, k, a) \, \sim \,
\, { < \int \sqrt{g} \, R > \over < \int \sqrt{g} > } ,
\eeq
and the fluctuation in the local curvatures
\beq
\chi_\cR  (\lambda, k, a) \, \sim \,
{  < ( \int \sqrt{g} \, R )^2 > - < \int \sqrt{g} \, R >^2
\over < \int \sqrt{g} > }
\sim  \frac{1}{V} \frac{\partial^2}{\partial k^2} \ln Z .
\eeq
The lattice analogs of these expressions are readily written down
by making use of the correspondences \cite{hw84,lesh}
\bea
\int d^4 x \, \sqrt{g} & \to & \sum_{\rm hinges \, h} V_h  \\
\int d^4 x \, \sqrt{g} \, R & \to & 2 \sum_{\rm hinges \, h} \delta_h A_h \\
\int d^4 x \, \sqrt{g} \, R_{\mu\nu\rho\sigma} R^{\mu\nu\rho\sigma}
& \to & 4 \sum_{\rm hinges \, h} V_h \, ( \delta_h^2 A_h^2 / V_h^2 ) .
\eea
On the lattice we prefer to define quantities in such a way that
variations in the average lattice spacing $\sqrt{<l^2>}$ are compensated by
the appropriate factor as determined from dimensional considerations.
In the case of the average curvature we define therefore the lattice quantity
$\cR$ as
\beq
{\cal R} \; = \; <l^2> { < 2 \sum_h \delta_h A_h > \over < \sum_h V_h > } ,
\label{eq:avr}
\eeq
and similarly for the curvature fluctuation.
The curvature fluctuation is related to the (connected) scalar curvature
correlator at zero momentum
\beq
\chi_\cR \sim { \int d^4 x \int d^4 y < \sqrt{g} R (x) \sqrt{g} R (y) >_c
\over < \int d^4 x \sqrt{g} > } .
\eeq
A divergence in the fluctuation is then indicative of
long range correlations (a massless particle).
Close to the critical point one expects for large separations a power
law decay in the geodesic distance,
\beq
< \sqrt{g} R (x) \sqrt{g} R (y) >
\mathrel{\mathop\sim_{ \vert x - y \vert \rightarrow \infty}}
\frac{1}{ \vert x-y \vert^{2n} } ,
\eeq
which in turn leads to the expectation $ \chi_\cR \sim L^{d-2n}$, where
$L \sim V^{ 1 / d} $ is the linear size of the system. In \cite{tala,ph1}
it was found that $\chi_\cR $ diverges close to the critical point as
\beq
\chi_\cR
\mathrel{\mathop\sim_{ k=k_c, \, L \rightarrow \infty}}
L^{ d ( 1 - \delta ) / ( 1 + \delta ) } ,
\eeq
where $\delta $ is the curvature critical exponent introduced in \cite{tala},
and therefore $ n = \delta d /( 1 + \delta ) = d - 1 / \nu  $, with
the exponent $\nu $ defined as $\nu = (1 + \delta)/d $.
Note that for a {\it scalar} field in four dimensions
one would expect $\nu = 1/2$, whereas we find
$\delta \approx 0.63$ and therefore $\nu \approx 0.41 $.

It is of interest to contrast the behavior of the preceding quantities,
associated with the curvature, with the analogous quantities involving
the local volumes (or the square root of the determinant of the metric in
the continuum) only.
We can consider therefore the average volume $<V>$, and its
fluctuation defined as
\beq
\chi_{V}  (\lambda, k, a) \, \sim \,
{  < ( \int \sqrt{g} )^2 > - < \int \sqrt{g} >^2
\over < \int \sqrt{g} > }
\sim  \frac{1}{V} \frac{\partial^2}{\partial \lambda^2} \ln Z .
\eeq
The latter is then related to the connected volume correlator at zero momentum
\beq
\chi_V
\sim { \int d^4 x \int d^4 y < \sqrt{g(x)} \sqrt{g(y)} >_c
\over < \int d^4 x \sqrt{g} > } .
\eeq
We have argued before \cite{tala}
that fluctuations in the curvature are sensitive
to the presence of a spin two massless particle,
while fluctuations in the volume probe only the correlations in the
scalar channel.
In the case of gravity a dramatic difference is therefore expected in
the two type of correlations.
Indeed the numerical simulations show clearly
a divergence in the curvature fluctuations, but at the same
time no divergence in the volume fluctuations.
Other, more complex invariant correlation functions at fixed geodesic
distance can be written down and measured \cite{ph1}.

Let us turn now to the observables involving the scalar field.
Due to the form of the action, the average of the scalar field
is always zero, but one can compute the discrete analog of the
following coordinate invariant fluctuation
\bea
\chi_\phi = &&
{ < \int d^4 x \int d^4 y \sqrt{g(x)} \, \phi (x) \sqrt{g(y)} \, \phi (y) >
\over < \int d^4 x \sqrt{g(x)} > }
\nonumber \\ &&
- \, { < \int d^4 x \sqrt{g(x)} \, \phi (x) >
< \int d^4 y \sqrt{g(y)} \, \phi (y) >
\over < \int d^4 x \sqrt{g(x)} > }
\eea
(again, for the Gaussian scalar action we will be considering, the
second term on the r.h.s. will be zero).
On the lattice such an expression can be written as
\beq
\chi_\phi \sim
{ < \sum_{i j} V_i \phi_i V_j \phi_j >
\over < \sum_i V_i > } -
{ < \sum_i V_i \phi_i > < \sum_j V_j \phi_j >
\over < \sum_i V_i > } .
\label{eq:phichi}
\eeq
Since $\chi_\phi $ is the zero-momentum component of the scalar
particle propagator, it is expected to diverge like $m^{-2}$ for small mass,
up to anomalous dimensions.
Also of interest are the local coordinate invariant averages
\bea
< \phi^2 > & \equiv &
{ < \int d^4 x \sqrt{g} \, \phi^2 > \over < \int d^4 x \sqrt{g} > }
\nonumber \\
< \phi^4 > & \equiv &
{ < \int d^4 x \sqrt{g} \, \phi^4 > \over < \int d^4 x \sqrt{g} > } .
\label{eq:phi2}
\eea
For {\it free} fields one expects the following dependence on the scalar
field mass,
\beq
< \phi^2 > \, = \int {d^4k \over ( 2 \pi )^4 }
\frac{1}{ k^2 + m^2 } =
\frac{1}{16 \pi^2} \left [ \Lambda^2 - m^2 \ln \frac{\Lambda^2 + m^2}{m^2}
\right ] ,
\label{eq:freephi2}
\eeq
\beq
< \phi^4 > \, = 2 \int {d^4k \over ( 2 \pi )^4 }
\frac{1}{ (k^2 + m^2)^2 } =
\frac{1}{8 \pi^2} \left [ \ln \frac{\Lambda^2 + m^2}{m^2} +
\frac{m^2}{\Lambda^2 + m^2} - 1 \right ] ,
\label{eq:freephi4}
\eeq
where $\Lambda$ is the ultraviolet momentum cutoff. In the interacting case
one anticipates, among other effects, a multiplicative renormalization
of the mass parameter $m$. In the presence of gravity, the behavior
of these quantities will be discussed below.

We can write schematically the propagator for the scalar field in a
{\it fixed} background geometry specified by some distribution
of edge lengths as
\beq
G(d) \, = \, < y | { 1 \over  - \partial^2 + \xi R + m^2 } | x > ,
\eeq
where $d$ is the geodesic distance between the two spacetime points being
considered.
Now fix one point at the origin $0$, and use the discretized
form of the scalar field action of Eq.~(\ref{eq:acp}). Then the discrete
equation of motion for the field $\phi_i$ in the presence of a
$\delta$-function source of unit strength localized at the origin
gives us the sought-after Green's function.
For $\xi=0$ we write the equation as
\beq
\phi_i = \frac{1}{W_i} \, \Bigl ( \sum_{ j \not= i }
W_{ij}  \; \phi_j + \delta_{i0} \Bigr ) ,
\eeq
with the weights $W$ given by
\beq
W_i = \sum_{j \not= i } \, \Bigl ( \frac{m^2}{2} + \frac{1}{l_{ij}^2}
\Bigr ) \, V_{ij} \;\;\;\;\;\;\; W_{ij} = \frac{V_{ij}}{l_{ij}^2} .
\eeq
Here the sums extend over nearest-neighbor points only,
$V_{ij}$ is the volume associated via a baricentric subdivision
with the edge $ij$, and
$\delta_{i0} $ is a delta-function source localized at the origin
on site $0$.
The above equation for $\phi_i$ can then be solved by an iterative procedure,
taking $\phi_i = 0$ as an initial guess.
After the solution $\phi_i$ has been
determined by relaxation, at large distances from the origin one has
\beq
\phi_i \sim G(d_{i0}) \sim A \, \sqrt{m / d_{i0}^{\, 3} }  \, \,
exp \, ( - m d_{i0} ) ,
\eeq
which determines the geodesic distance $d_{i0}$ from lattice point $0$ to
lattice point $i$. This method is more efficient
and accurate than trying to determine the geodesic distance by sampling paths
connecting the two points as was done in \cite{ph1},
but is of course equivalent to it \cite{fi}.

In quantum gravity it is of great interest to try to determine the
value of the low energy, renormalized coupling constants, and in
particular the effective cosmological constant $\lambda_{eff}$ and the
effective Newton's constant $G_{eff} = (8 \pi k_{eff})^{-1}$.
Equivalently, one would like
to be able to determine the large distance limiting value of a
dimensionless ratio such as $\lambda_{eff} G_{eff}^2 $, and its
dependence on the linear size of the system $L = V^{1/4} $.
(In the real world one knows that
$G_{eff} = (1.6160 \times 10^{-33} cm )^2 $, while
$\lambda_{eff} G_{eff}^2 \sim 10^{-120} $ is very small).
The vacuum expectation value of the scalar curvature can be used as
a definition of the effective, long distance cosmological constant
\beq
{\cal R} \, \sim \, { < \int \sqrt{g} \, R > \over < \int \sqrt{g} > }
\sim \left ( { 4 \lambda \over k } \right )_{eff} .
\label{eq:effr}
\eeq
In the pure gravity case one finds that there is a critical point in
$k$ at which the curvature vanishes, and for $k < k_c$ one has
\beq
\cR
\mathrel{\mathop\sim_{ k \rightarrow k_c}}
- A_\cR \, ( k_c - k )^\delta
\eeq
and thus $ (\lambda / k )_{eff} \rightarrow 0 $ in lattice units.
The location of the critical point $k_c$ and the amplitude in general depend on
the higher derivative coupling $a$ and other non-universal parameters,
but the exponent is expected
to be universal, and was estimated previously to be about 0.63; more
details can be found in refs. ~\cite{tala,ph1}.

One immediate consequence of this result is that in the smooth phase
with $k < k_c$ (or $G > G_c \equiv G^*$) the gravitational coupling
constant $G$ must increase with distance (anti-screening), at least
for rather short distances.
Introducing an arbitrary momentum scale $\mu $, one has close to the
ultraviolet fixed point the following short-distance behavior for
Newton's constant
\beq
G ( \mu ) - G^* \, = \, \left [  G ( \Lambda ) - G^* \right ]
\left ( \Lambda \over \mu \right ) ^{ 1 / \nu }
\eeq
with $\Lambda $ the ultraviolet cutoff; the exponents $\delta$ and $\nu$
are calculable and are related to each other via the scaling relation
$\nu = (1 + \delta ) / 4 \approx 0.41$.
The opposite behavior (screening) would be true in the phase with
$k > k_c$, but such a phase is known not to be stable and leads to no lattice
continuum limit \cite{ph1}.

If the system is of finite extent, with linear dimensions $L=V^{1/4}$,
then the scaling laws for ${\cal R} $ should also give the volume
dependence of the effective cosmological constant at the fixed point.
For pure gravity one finds at the critical point
\beq
{\cal R} \sim \, \mathrel{\mathop\sim_{ L \gg l_0 }} \,
\left ( { 1 \over L } \right )^{ \delta / \nu} ,
\label{eq:rsize}
\eeq
with $l_0$ of the order of the average lattice spacing,
$l_0 = \sqrt{ < l^2 > } $, and $\delta / \nu \approx 1.54 $.
The critical point here is defined, as usual, as the point in bare
coupling constant space where the curvature fluctuations diverge in the
infinite volume limit. Similarly for the dimensionless coupling $G$ in a finite
volume, one expects the scaling behavior
\beq
G ( \mu )  \, \mathrel{\mathop\sim_{ L, \; 1 / \mu \gg l_0 }} \,
G_c + \left ( { 1 \over \mu L } \right )^{ 1 / \nu} ,
\label{eq:gsize}
\eeq
These results are all direct consequences of the scaling laws
and the values of the critical exponents \cite{ph1}.
An important issue is how these results are affected by the presence
of dynamical matter. This will be addressed later in the paper.

The gravitational exponent $\delta$ determines the universal scaling
behavior of a variety of observables. Among the simplest ones which
are relevant for simple cosmological models one can mention the FRW
scale factor $a(t)$, as it appears in the line element
\beq
d s^2 \, = \, - dt^2 + a^2 (t) \left \{ { dr^2 \over 1 - k r^2 } +
r^2 ( d \theta^2 + \sin^2 \theta d \phi^2 ) \right \} ,
\eeq
and which we would expect to scale at short distances according to the equation
\beq
{ a^2 (t) \over a^2 (t_0) } \; \mathrel{\mathop\sim_{ t \gg t_0 }} \;
\left ( { t \over t_0 } \right )^{ \delta / \nu }
\eeq
with $c t_0 = l_0 $.
It is amusing to note that in this model the scale factor cannot exhibit a
singularity for short times, $t \sim t_0$. For such short distances the
strong fluctuations in the metric field and the curvature
prevent this from happening. We should add though that
the scale factor itself is essentially a semiclassical quantity,
linked to a specific ansatz for the (classical) metric at large
distances. In the presence of strong metric fluctuations it is no
longer clear that it remains a well-defined concept.

The bare Newton's constant also describes the coupling of gravity to
matter at scales comparable to the ultraviolet cutoff.
Consider the classical equations of motion
for pure Einstein gravity with a cosmological constant term
\beq
R_{\mu\nu} - \half g_{\mu\nu} R + \Lambda g_{\mu\nu} = 8 \pi G T_{\mu\nu} .
\eeq
Here we have followed the usual conventions by defining
$ \Lambda = 8 \pi G \lambda $ (not to be confused with the ultraviolet
momentum cutoff introduced earlier).
In the presence of higher derivative terms and higher order lattice
corrections this is of course not the right equation (the equations of
motion for higher derivative gravity are substantially more complex),
but at sufficiently large distances it should be the appropriate
equation if the average curvature is small and a sensible continuum
limit can be found in the lattice theory.
If we have only one real scalar field, the energy-momentum tensor is given by
\beq
T_{\mu\nu} =  \partial_\mu \phi \, \partial_\nu \phi -
\half g_{\mu\nu} ( \partial_\lambda \phi \, \partial^\lambda \phi
+ m^2 \phi^2 )
\eeq
(we will consider from now on only the case $\xi=0$).
Taking the trace we obtain
\beq
R = 4 \Lambda - 8 \pi G \, T_\mu^\mu = 4 \Lambda + 8 \pi G \,
\left [ (\partial \phi )^2 + 2 m^2  \phi^2 \right ] .
\eeq
Now consider the effects of quantum fluctuations, and
separate the pure gravity and matter contributions to the scalar curvature,
by writing for the average curvature
$ < R > \, = \, < R_{gravity} > + < R_{matter} > $,
where  $ < R > $ is the average of the
total scalar curvature in the presence of matter,
and $ < R_{gravity} > $ is the same quantity in the absence of matter.
More specifically, by the expectation value $ < R_{gravity} > $
we will simply mean the averages obtained in the absence of
any matter fields, as computed in ref. \cite{ph1}.
We will see below that $ < R_{matter} > $ represents a rather small
contribution, unless there are many scalar fields contributing
to the vacuum polarization.
In the presence of quantum fluctuations, we can write therefore for the
matter correction
\beq
< R_{matter} > \, = 8 \pi G
< (\partial \phi )^2 + 2 m^2  \phi^2 >
\, = \, 8 \pi G \left [ 2 < I_{\phi} > + m^2  < \phi^2 > \right ] .
\label{eq:geff}
\eeq
In other words, the change in the average value of the
scalar curvature that arises when matter fields are included is
proportional to Newton's constant $G$, and it is expected to
be positive.
This is indeed what will be found in the numerical simulations
discussed
below, even though the magnitude of the correction is quite small
(in agreement with the perturbative arguments presented in the
previous section).
To the extent that the feedback of the scalar degrees of freedom
on the gravitational degrees of freedom appears to be rather small (almost
to the point of being difficult to measure), we shall
argue below that gravity is indeed 'weak', at least for the
type of scalar action we have investigated here.

\vskip 10pt
\newsection{Numerical Procedure}

In order to explore the ground state of four-dimensional simplicial
gravity coupled to matter beyond perturbation theory one has to resort
to numerical methods.
As in our previous work, the edge lengths and scalars are updated by a
standard Metropolis algorithm,
generating eventually an ensemble of configurations distributed
according to the action of Eqs.~(\ref{eq:acg}) and ~(\ref{eq:acp}),
with the inclusion
of the appropriate generalized triangle inequality constraints
arising from the nontrivial gravitational measure.
Further details of the method as applied to pure gravity
are discussed in \cite{ijsa}, and will not be repeated here, since
the scalar action contribution can be dealt with in essentially the same way.

We have not included here a term coupling the scalar field directly to the
curvature ($\xi =0$), since the continuum perturbative results discussed
previously appear rather similar for different values of $\xi \ne \sixth $,
and the scalar action becomes significantly simpler for $\xi =0$. Also we note
that, in the absence of matter,
$<R>$ itself vanishes at the critical point \cite{tala,ph1}.
In mean field theory, we can replace the term $R \phi^2 $ by $R < \phi^2 > $.
Since $< \phi^2 > $ is finite at the critical point (see discussion
below), we expect the inclusion of this term to mostly affect a
renormalization of the critical coupling $k_c$ (related to the
critical value of Newton's constant by $k_c = 1 / (8 \pi G_c)$),
which should not change the universal critical behavior.

Let us point out here only the fact that, while the scalar field
action of Eq.~(\ref{eq:acp}) looks rather innocuous, due to the
simplicial nature
of the lattice a large number of interaction terms are involved at
each site: at each vertex there are 15 edges emanating in the positive
lattice 'directions', and 15 in the negative lattice 'directions' \cite{rw}.
In the update of the scalar field each of the 30 edge volumes
$V_{ij}$ has to be re-computed, by adding together the contributions
from all the four-simplices that meet on that edge. For the edge
volume one has
\beq
V_{ij} \, = \, \tenth \sum_{\rm simplices \,\, s \supset ij } V_s
\eeq
since there are ten edges per simplex in four dimensions.
Here the volume of a n-simplex with edge lengths $l_{ij}$
is given as usual by the determinant
\beq
V_n = { (-1)^{n+1 \over 2} \over n!  2^{n/2} }
\left| \begin{array}{llll}
0      &    1     &    1     & \ldots \\
1      &    0     & l_{12}^2 & \ldots \\
1      & l_{21}^2 &    0     & \ldots \\
1      & l_{31}^2 & l_{32}^2 & \ldots \\
\ldots &  \ldots  &  \ldots  & \ldots \\
1      & l_{n1}^2 & l_{n2}^2 & \ldots \\
1      & l_{n+1,1}^2 & l_{n+1,2}^2 & \ldots \\
\end{array} \right| ^{1 \over 2} , \\
\eeq
and corresponds to the determinant of a $6 \times 6$ matrix in the case
of a four-simplex; when expanded out it contains 130 distinct terms.
Furthermore the number of four-simplices meeting on a given edge depends
on the type of edge. With our simplicial subdivision of the four-dimensional
hypercubes that make up the lattice, we have four body principals,
six face diagonals, four body
diagonals and one hyperbody diagonal per hypercube \cite{rw}.
For a body principal or hyperbody diagonal
there are 24 four-simplices meeting on it, while for a face or body
diagonal there are 12 four-simplices meeting on it.
When updating one scalar field by the multi-hit Monte Carlo or heat bath
method, the 30 neighboring
link contributions need to be computed once, with their associated link
volumes, and special care has to be taken of the order of the edge lengths
appearing in the simplex formulae.
When updating a given edge length, all the scalar field action
contributions involving that particular edge have to evaluated, in addition
to the purely gravitational part. For a body principal and hyperbody
diagonal there are 65 such contributions that have to be added up,
while for a face or body diagonal 35 such contributions have to be added up.
By assigning then special fixed values to the edge lengths, one can perform
a number of checks against the expected analytical result to verify that
the volumes are computed and added up correctly.
Even though the program is quite computing intensive, it is well suited
for a massively parallel machine. In the two parallel versions of the
program we have written, a large number (64-256) of independent edge
and scalar variables are all updated simultaneously in parallel.

We considered lattices of size between $4 \times 4 \times 4 \times 4$
(256 vertices, 3840 edges, 6144 simplices)
and $16 \times 16 \times 16 \times 16 $
(65536 vertices, 983040 edges, 1572864 simplices).
Even though these lattices are not very large, one should keep in mind
that due to the simplicial nature of the lattice there are many edges
per hypercube with many interaction terms, and as a consequence the
statistical fluctuations are comparatively small, unless one is very
close to a critical point.
In all cases the measure over the edge lengths was of the form
$ dl^2 V_l^{n_f/30} $ times the triangle inequality constraints
(see Eq.~(\ref{eq:meas})). We shall restrict here our attention to
the case $n_f=1$; results for larger values of $n_f$ will
be presented elsewhere.

The topology was restricted to a four-torus (periodic boundary conditions),
and it is expected that for this choice boundary effects on physical
observables should be minimized.
One could perform similar calculations with lattices of different
topology, but the universal infrared scaling properties of
the theory should be determined only by short-distance renormalization
effects, independently of the specific choice of boundary conditions.
This is a consequence of the fact that the renormalization group
equations are {\it independent} of the boundary conditions, which enter only
in their solution as it affects the correlation functions through
the presence of a new dimensionful parameter $L$.
Thus the four-torus should be as good as any other choice of topology,
as long as we consider the universal long distance properties.

Let us give here a few details about the runs performed to compute
the averages.
In the presence of matter fields, the
lengths of the runs are much shorter than in the pure gravity
case \cite{ph1}, since the scalar field update is rather time-consuming.
The couplings $\lambda$ and $a$ in the gravitational action
of Eq. ~(\ref{eq:acg}) were fixed, as in the pure gravity case, to
$1$ and $0.005$ respectively. For pure gravity this choice leads
to a well defined ground state for $k \le k_c \approx 0.244 $ (the system
then resides in the smooth phase, with a fractal dimension
very close to four).
In the presence of matter, we also restricted most of our runs to this
physically more interesting phase, where the curvature is small and negative.
We investigated five values of $k$ ($0.0,0.05,0.1,0.15,0.20$), and for
each value we looked at a scalar mass of $1.0$, $0.5$ and $0.2$ in
lattice units.
In addition, we have accurate results for infinite mass from the previous
{\it pure} gravity calculations. Besides the results on lattices with
$L=4$ for all the above values of $k$ and $m$,
we also have accurate results on lattices of size $L=8$ and 16 for
$m=0.5$, and of size $L=8$ for $m=0.2$.
For these values of the scalar mass, the scalar correlations only extend over
a few lattice spacings, and finite size effects should therefore be contained
(we have checked that this is indeed the case for the quantities we
have measured).
In general we are interested in a regime in which the scalar mass
is much larger than the infrared cutoff, but much smaller than
the lattice ultraviolet cutoff, or
\beq
\sqrt{< l^2 >} \; \ll \;  m^{-1} \; \ll \; V^{1/4} ,
\eeq
in order to avoid finite lattice spacing and finite volume effects.
Similarly, one should also impose the constraint that
the scale of the curvature in magnitude should be much smaller than
the average lattice spacing, but much larger than the size of the
system, or
\beq
< l^2 >  \; \ll \;\; < l^2 > | {\cal R} |^{-1} \; \ll \; V^{1/2} .
\eeq
It is equivalent to the statement that in momentum space the physical scales
should be much smaller that the ultraviolet cutoff, but much larger
than the infrared one.

The lengths of the runs typically varied between $2-6k$ Monte Carlo
iterations on the $4^4$ lattice, $1-2k$ on the $8^4$ lattice, and
$0.6-0.9k$ on the $16^4$ lattice.
The runs are comparatively longer on the larger lattices, since it
was possible in that case to use a fully parallel version of the program.
As input configurations, we used the thoroughly thermalized configurations
generated previously for pure gravity. These configurations
are rather 'close' to the ones that include the effects of matter, since
the feedback of matter turns out to be rather small.
On the larger lattices duplicated copies of the
smaller lattices are used as starting configurations
for each $k$, allowing for additional
equilibration sweeps after duplicating the lattice in all four directions.
This allows for a substantial savings in time, since the initial
edge length configuration on the larger lattice is already quite close to
a representative configuration. We have found that in the well behaved
phase ($k < k_c$) the auto-correlation times are contained, of the
order of at most about one hundred sweeps. When we duplicate the
smaller lattice to a larger lattice, almost no drift in the averages is
observed during later re-thermalization, which indicates that for
our parameters the finite size corrections are small. On the larger
lattices, because there are so many variables to average over, the
statistical fluctuations from configuration to configuration are of
course much smaller.

\vskip 10pt
\newsection{Results}

In the pure gravity case, one finds that for fixed positive $a$ and $\lambda$
(the latter can be set equal to one without loss of generality, since
it determines the overall scale)
and sufficiently small $k$, the curvature is small and negative (smooth
phase), and goes to zero at the critical point $k_c (a) $, where
the curvature fluctuation diverges.
In the pure gravity case we write therefore, for $k$ less than $k_c$
\beq
\cR (k, a)
\mathrel{\mathop\sim_{k \rightarrow k_c(a)}}
- A_\cR (a) \, \left ( k_c(a) - k \right )^\delta
\label{eq:rsing}
\eeq
\beq
\chi_\cR (k, a)
\mathrel{\mathop\sim_{k \rightarrow k_c(a)}}
A_\chi (a) \, \left ( k_c(a) - k \right )^{\delta-1}
\eeq
where $\delta$ is a universal curvature critical
exponent, characteristic of the gravitational
transition \cite{tala}.
Here we will only consider the case $a=0.005$, for which the phase transition
is second order, leading therefore to a well-defined continuum limit
at least in the pure gravity case \cite{ph1}.
For $k \ge k_c$ the curvature is very large (rough phase),
and the lattice tends to collapse into degenerate configurations
with very long, elongated simplices ( with $ <V_h> / <l^2>^2 \sim 0$).
(In ref. \cite{ph1} several values for $a$ were studied, and it was found
that the model actually exhibits multicritical behavior.
While for $a=0.005$ one finds a second order phase transition,
for $a=0$ the singularity
appears to be in fact logarithmic ($\delta=0$), suggesting
a first order transition with no continuum limit for sufficiently
small $a$, with a multicritical point separating the two transition
lines).

When including the effects of the scalar field, one finds that the largest
changes are in the average volumes (which decrease by about three
percent for a scalar mass $m=0.5$)
and the average edge lengths. But such changes
are somewhat uninteresting, since they correspond effectively to a shift
(here actually an increase) in the bare cosmological constant (also by
about the same percentage, since
$\delta V / V \sim - \delta \lambda / \lambda$).
We note here incidentally that such a small effect is consistent with
the perturbative result of
Eq. (\ref{eq:effcoupl}), which predicts an increase in the effective
cosmological constant $\lambda$ by about one percent, for a cutoff
$\Lambda \sim \pi / l_0 \sim 1 $.
Indeed before we have chosen to define observables in such
a way that these effects are largely compensated, by rescaling by an
appropriate power of the average lattice spacing, as in Eq. (\ref{eq:avr}).
Physically more interesting are the results for the average curvature
in the presence of the scalar field.
As can be seen from Fig. 1, the effects of the feedback of one scalar
field on the curvature are quite small. It is useful to display the results
as a function of $ z = 1 / (1 + m^2 )$, since this allows us to put the
results for infinite mass (no scalar feedback, from ref. \cite{ph1})
on the same graph. The most accurate results in the presence of the
scalar field are for $m=0.5$, where we have relatively accurate results
for three different lattice sizes ($L=4,8,16$) and the highest statistics.
The points for $m=1.0$ are for reference only, since they
are from an $L=4$ lattice only.
For $m=0.5$ and $m=0.2$ the results show a small but clear systematic
decrease in the magnitude of the average curvature in the smooth phase for all
values of $k$, at the level of a few percent; to see such a small
effect long runs were needed.
The results are in qualitative agreement with the expectation that
the presence of the scalar field should give a positive
contribution to the average curvature.
In any case, for all values of the mass we have considered, the effects are
rather small.

As should be clear from the discussion in the previous section, we are
interested in how the critical behavior of the theory is affected in the
neighborhood of the critical point by the presence of the scalar field.
We will write therefore again for the average curvature, now in the
presence of the scalar field,
\beq
\cR  \mathrel{\mathop\sim_{ k \rightarrow k_c }}
- A_{\cR} \, ( k_c - k )^\delta ,
\eeq
where now we expect $A_{\cR} , k_c , \delta $ to depend also on the number of
scalar flavors, $n_f$.
In the presence of the scalars we have to look at the scaling limit
$m \rightarrow 0$, which in practical terms corresponds to a mass much
smaller than the inverse average lattice spacing.
It is not clear if $m=0.5$ (where we have our most accurate results)
in our case corresponds already to such a
scaling region, but our results should not be too far off, if the experience
in other lattice models can be used here as a guide.
If we adopt the same procedure as for pure gravity, and fit the average
curvature for $m=0.5$ to an algebraic singularity, we find
$A_{\cR}=3.68(5) $, $k_c=0.243(2)$ and $\delta=0.61(6)$.
This should be compared to the estimates for pure gravity (and for the same
value of $a=0.005$),
$A_{\cR}=3.79(4) $, $k_c=0.244(1)$ and $\delta=0.63(3)$ \cite{ph1}.
In Fig. 2 we compare the results for the average curvature $\cR(k)$ with
and without the presence of the scalar fields.
In Fig. 3 the same data is used to display $ [ - \cR (k) ]^{1 / \delta} $
instead, which as can be seen from the graph
deviates very little from a straight line behavior in $k$, if
one uses $\delta = 0.63 $.

We conclude therefore that, within our errors, switching on the
scalar fields leaves the exponents almost unchanged, and the critical
point moves very little; our results suggests that
$k_c$ decreases when we include the effects of the scalar field.
Again we notice that such a small shift is not unexpected on the basis
of the perturbative result of Eq. (\ref{eq:effcoupl}), which also
suggests a small decrease in the effective $k$, for a cutoff
$\Lambda \sim \pi / l_0 \sim 1 $.
For small non-integer $n_f$ we can expand the amplitude, critical
value of $k$ and the exponent in powers of the number of flavors $n_f$,
\bea
& A_{\cR} & =  A_0 + n_f A_1 + O(n_f^2) \nonumber \\
& k_c & =  k_0 + n_f k_1 + O(n_f^2) \nonumber \\
& \delta & =  \delta_0 + n_f \delta_1 + O(n_f^2) ,
\eea
and for the average curvature itself we get
\beq
\cR \mathrel{\mathop\sim_{ n_f \rightarrow 0 }}
- A_0 ( k_0 - k )^{\delta_0} \left \{
1 + n_f \bigl [ \frac{A_1}{A_0} + \frac{\delta_0 k_1}{k_0-k}
+ \delta_1 \ln (k_0 -k) \bigr ] + O(n_f^2) \right \}  ,
\eeq
which shows that the $k_1$ renormalization is dominant for very small
$n_f$.
Since the results for $n_f=1$ indicate that the corrections due to the scalar
field are quite small, we would tend to conclude that coefficients
of the $n_f$ terms must be rather small, and that the pure gravity theory
is already a good approximation to the full theory including scalars,
provided $n_f$ is not too large.

Let us assume for the moment that $k_1$ and $\delta_1$ are so small that
they can be neglected to a first approximation when we consider a single
scalar matter field (in the $2+\epsilon$ expansion the matter corrections
are certainly very small, and the exponent is independent of the number
of matter fields to leading order in $\epsilon$).
Then the difference between the average curvature in the presence of the
scalar field and in pure gravity determines the ratio of curvature
amplitudes $A_1/A_0$,
\beq
{ \cR_{matter} \over \cR_{gravity} } \, = \,
{ \cR_{gravity+matter} - \cR_{gravity} \over \cR_{gravity} }
\mathrel{\mathop\sim_{ k \rightarrow k_c }}
{ A_1 \over A_0 }
\eeq
The difference in the numerator is of course quite small, and requires
a very accurate measurement of the average curvature in both cases.
At the same time it
provides a direct determination of the physical effects of dynamical
matter fields, on a quantity that represents a direct physical observable,
since the average curvature can in principle be measured by performing
parallel transports of vectors around large closed loops.
The calculated difference
$ \cR_{gravity+matter} - \cR_{gravity} $ is shown in Fig. 4, together
with a fit to a behavior $ \sim ( k_c - k )^{\delta} $, treating only the
amplitude as a free parameter.
To reduce any systematic effects coming from finite
volume corrections, it is advisable to subtract the
average curvatures on the {\it same} lattice size.
In addition, such a subtraction can be done without any assumption
about the (singular) behavior of the curvature at $k_c$.
One then estimates approximately for the ratio
$ A_1 / A_0 \approx 0.053 / 3.79 = 0.014 $;
we will leave a more accurate quantitative determination of this ratio
for future work. We note though that the sign of the matter correction
to the curvature is consistent with the fact that the effective
Newton's constant gives rise to an attractive interaction ($G_{eff} >
0)$, thereby adding a positive contribution to the pure gravity
average curvature.

For an explanation for the smallness of such a ratio, we can look again
at the formula of Eq. ~(\ref{eq:nh}).
There the relative smallness of the matter
contribution is simply a consequence of the particle's relative spin.
For spin zero and spin two, as we have here, the ratio of the matter
over gravity contributions is
$ \sim \third / ( 4 s^2 - \third ) = 0.021 $, indeed of the same order as the
ratio we computed. One can go perhaps as far as turning
this argument around, and argue that
the smallness of the vacuum polarization effects compared to the
purely gravitational contribution is an indirect indication of the spin-two
nature of the graviton (if we were to treat the value of the graviton
spin as an unknown parameter, we would obtain a value
very close to two, $s \sim 2.5 $).

Let us turn now to a discussion of the renormalization properties of
the couplings $ G$ and $\lambda$.
It is clear from the preceding discussion that the effects of scalar
matter are quite small. In the following we shall therefore not
distinguish between the cases with and without matter fields, assuming
that if there are only a few matter fields, the exponents will not
change drastically.

As we indicated previously,
using the methods of finite size scaling \cite{fss},
one can translate the dependence of the curvature on $k-k_c$
into a statement about the {\it volume}
dependence of the curvature at the critical point $k_c$.
In a finite volume, of linear size $L$, finite size scaling
(from Eqs. ~(\ref{eq:effr}) and ~(\ref{eq:rsize})) gives
\beq
( G \lambda )_{eff} ( L ) \,
\mathrel{\mathop\sim_{ L  \gg l_0 }} \,
l_0^{-2} \left ( { l_0 \over L } \right )^{ 4 - 1 / \nu} ,
\eeq
since essentially the correlation length $\xi$ saturates at the system size,
$ \xi \sim (k_c - k)^{-\nu} \sim L $.
Combining this result with Eq. ~(\ref{eq:gsize}), one obtains for
the dimensionful Newton's constant the following scale dependence,
valid for short distances $1 / \mu \ll L $,
\beq
G_{eff} ( \mu ) \, \mathrel{\mathop\sim_{ L , \; 1 / \mu \gg l_0 }} \,
l_0^{2} G_c + l_0^{2} \left ( { 1 \over \mu L } \right )^{ 1 / \nu} ,
\eeq
(with $ 1 / \nu \approx 2.46 $), and for the dimensionful cosmological constant
\beq
\lambda_{eff} ( \mu ) \, \mathrel{\mathop\sim_{ L , \; 1 / \mu \gg l_0 }} \,
l_0^{-4} \left ( { \mu l_0 } \right )^{ 4 - 1 / \nu}
\left [ G_c + \left ( { 1 \over \mu L } \right )^{ 1 / \nu} \right ]^{-1}
\eeq
(with $ 4 - 1 / \nu \approx 1.54 $),
Here again $l_0$ is of the order of the average lattice spacing, and we have
restored the correct dimensions for $G_{eff}$ (length squared) and
$ \lambda_{eff} $ (inverse length to the fourth power).
For the dimensionless ratio $ G^2 \lambda$ we then obtain the result
\beq
( G^ 2\lambda )_{eff} ( \mu ) \,
\mathrel{\mathop\sim_{ L , \; 1 / \mu \gg l_0 }} \,
\left ( { \mu l_0 } \right )^{ 4 - 1 / \nu}
\left [ G_c + \left ( { 1 \over \mu L } \right )^{ 1 / \nu} \right ]
\eeq
As a check, it is immediate to see that the exponent associated with
$G_{eff}$ is indeed what one would expect from the form of the
Einstein part of the gravitational action in Eq. ~(\ref{eq:acgc}) and
the value of the curvature critical exponent $\delta$,
irrespective of whether matter fields are present or not (the specific
values of $\delta$ and $\nu $ will depend of course on how many matter
fields are present).

In conclusion, it seems that the dimensionless ratio $G^2 \lambda$
can be made very
small, provided the momentum scale $\mu$ is small enough, or, in other
words, at sufficiently large distances.
We should add also that the fixed point value for the dimensionless
gravitational constant, $G_c$, is in general non-universal and
cutoff-dependent, and depends on the specific way in which
an ultraviolet cutoff is introduced in the theory (here via an average
lattice spacing).
In our model it is of order one for very small $a$, but for larger $a$
it decreases in magnitude.

One notices that the smaller $G_c$, the smaller the distance
dependence of $G(r)$, since one has for the distance variation the result
\beq
{ \delta G ( r ) \over G ( r ) } \, = \,
{ \nu^{-1} \over G_c ( L / r )^{ 1 / \nu} + 1 } \, { \delta r \over r } ,
\eeq
(we have set $r= 1 / \mu$), so in practice $G_c$ cannot be too large.
For small $G_c$, $l_0^2$ becomes substantially larger than the Planck length.
It should be pointed out here that there is apparently no reason why
in this model the effective coupling $G_{eff}$ should turn out to be
of the same order as the ultraviolet cutoff $l_0^{-1}$, and indeed it
does not; the previous results seem to indicate that the situation
is more subtle.
Let us also add that one does do not expect the results to depend
significantly on the form of the lattice scalar action we have used.
In particular the presence of additional higher derivative terms
involving the scalar fields should not affect the results close to the
continuum limit, since the corrections should be suppressed by inverse
powers of the ultraviolet cutoff.

Another simple way of interpreting the results related to the
scalar field is as follows.
Close to the critical point, the average curvature approaches zero, and
at large distances it is therefore legitimate to write
$g_{\mu\nu} = \eta_{\mu\nu} + h_{\mu\nu}$, where $\eta_{\mu\nu}$ is the
flat metric, and $h_{\mu\nu} $ is a small correction. Then the scalar field
action of Eq.~(\ref{eq:acp}) is, again at large distances, close to the
action describing a free scalar, and its coupling to gravity is
correspondingly weak. At short distances the geometry fluctuates
wildly, and the coupling between gravity and matter is strong, while
at large distances the fluctuations eventually average out to zero,
effectively reducing the coupling.

Turning to the behavior of the scalar field itself, we show in Fig. 5
the results for $< \phi^2 >$,
in Fig. 6 those for $< \phi^4 >$ (see Eq.~(\ref{eq:phi2})), and
in Fig. 7 for $ \chi_{\phi} $ (defined in Eq.~(\ref{eq:phichi})).
The behavior of these three quantities is qualitatively rather
similar to their free field behavior (Eqs.~(\ref{eq:freephi2})
and ~(\ref{eq:freephi4})),
and is not too sensitive, at the
level of our accuracy, to the value of $k$. We note in particular that
$< \phi^2 > $ approaches a constant at $m=0$, while both $< \phi^4 > $ and
$\chi_{\phi}$ diverge at $m=0$, in agreement with a multiplicative mass
renormalization (no shift in the critical point for the field $\phi$, which
remains at $m=0$).


Let us conclude this section with a brief, qualitative discussion of
the phase diagram, reconsidered in light of the results obtained in
the presence of scalar matter.
In the case of pure gravity, the phase diagram shows a line of
critical points in the ($a,k$) plane separating the smooth from the
rough (or collapsed) phase of gravity.
The curvature vanishes along this line when it is approached from the
smooth phase, and for some sufficiently
negative $a < a_0 < 0 $ a stable ground state
ceases to exist entirely.
For $a=0$ or very small positive $a$, the transition from one phase
to the other is first order, with no continuum limit, while for larger
$a$ is becomes second order, with a well defined lattice continuum
limit, as we indicated previously. These findings in
particular would seem to indicate the presence of a multicritical point,
where the two transition lines intersect \cite{ph1}.

In the presence of scalar matter fields, and for sufficiently large $a$,
our new results presented here
seem to suggest that a continuum limit still exists.
In addition, we have found that in the smooth phase the average curvature
decreases in magnitude by a small but calculable relative amount.
A quantitative estimate for the amount of this decrease gives
$\Delta \cR / \cR \sim A_1 / A_0 \approx 0.014 $. As the number of
(degenerate) scalar fields increases, we expect this trend to continue, until
$\Delta \cR / \cR \sim n_f A_1 / A_0 \sim 1 $, at which point a
new phase transition might take place, in the sense that the
smooth phase disappears altogether (we expect that
the critical value $k_c$ will continue
to decrease, and might even become negative at some point).
The appearance of a new phase in the presence of matter,
with the geometry resembling
branched polymers, is a well known fact in two dimensions \cite{2dtrees}.
In Fig. 8 we have sketched what a possible phase diagram in the ($k,n_f$)
plane might look like.
Presumably this new phase is nothing but the rough phase found for
$n_f=0$ and sufficiently large $k$.
It is characterized by very long
elongated simplices, with very small volumes, and a fractal dimension
much smaller than four, reminiscent of a tree-like structure of space-time.
Given our rather limited results, a crude estimate
for the critical number of flavors at which this is expected to happen
would be $n_f \sim 71$, a rather large number.
But such an estimate is not inconsistent with the perturbative estimates
of Eqs.~(\ref{eq:beta}) and ~(\ref{eq:nh}), which also give such large
numbers (24 and 47, respectively).
And of course for such large values, we expect deviations
from linearity in $n_f$, and we will have to leave a direct investigation
of this issue for future work. Finally let us remark that since
the effects of fermions can be mimicked by having scalars with negative $n_f$,
the above conclusions would be rather different in that case, and their
presence should rather impede the appearance of this new phase transitions.
While scalars tend to make the geometry rougher, fermions should make
it smoother.

\vskip 10pt
\newsection{ Volume and Curvature Distributions }

In this section we will discuss the properties of volume and
curvature distributions, and how their behavior close to the critical
point, which defines the lattice continuum limit, can be related largely
to the critical exponents discussed previously.
Let us assume that close to the critical point $\lambda_c$ one has
for the average volume a singularity of the type
\beq
< V > \, \equiv \, < \int \sqrt{g} > \,
\sim - {\partial \over \partial \lambda } \ln Z
\mathrel{\mathop\sim_{ \lambda \rightarrow \lambda_c }}
{ V_0 \over ( \lambda - \lambda_c )^\omega } + {\rm reg.} ,
\eeq
with $\omega \neq 1 $, and ``${\rm reg.}$'' denotes the regular part.
For the volume fluctuation one then expects close to $\lambda_c$
\beq
< V^2 > - <V>^2 \, \sim \, {\partial^2 \over \partial \lambda^2 } \ln Z
\mathrel{\mathop\sim_{ \lambda \rightarrow \lambda_c }}
{ \omega V_0 \over ( \lambda - \lambda_c )^{\omega +1} } + {\rm reg.} ,
\eeq
and it follows that the partition function close to the singularity is
given by
\beq
Z_{sing.} ( \lambda ) \sim \exp \left \{ - \int^\lambda d \lambda'
{ V_0 \over ( \lambda' - \lambda_c )^\omega } + {\rm reg.} \right \} .
\eeq
Now let us introduce the quantity $N(V)$ defined by
\beq
N ( V ) = \int d \mu [g] \, \delta ( \int \sqrt{g} - V ) \, e^{-I[g]} .
\eeq
It can be evaluated from
\beq
N (V) = { 1 \over 2 \pi i } \int_{- i \infty }^{ + i \infty }
d \lambda \, Z ( \lambda ) \, e^{ \lambda V } ,
\eeq
to give, in the saddle point approximation, the following expression
for the density of states
\beq
N ( V ) \sim V^{ \gamma - 3 } \exp \left \{
\lambda_c V ( 1 + b / V^{1 / \omega } ) \right \} .
\eeq
where $b$ is a constant involving $\omega$, $V_0$ and
$\lambda_c$, and the exponent $\gamma$ parameterizes a possible power
law correction.
Let us denote by $<...>_V$ the averages obtained in the fixed volume
ensemble. Then it is easy to see, from the transformation properties of
the fixed-volume partition function under a change of scale, that one has
\beq
{ \partial \ln N(V) \over \partial V } =
- \frac{1}{V} + \frac{\sigma}{4}  + \frac{k}{2}
{ < \int \sqrt{g} R >_V \over V } ,
\eeq
which can be combined with the previous equation
to give the result, valid for large
volumes and in the fixed volume ensemble \cite{tala},
\beq
{ < \int \sqrt{g} R >_V \over V }
\mathrel{\mathop\sim_{ V \rightarrow \infty }}
c_0 - {  2 - \gamma \over V } + { c_1 \over V^{1 / \omega } } + \cdots .
\eeq
We have not calculated the above average in the fixed volume ensemble,
but in the {\it canonical} ensemble, where the volume is allowed to fluctuate,
one finds the following result close to the critical point \cite{ph1}
\beq
{ < \int \sqrt{g} R > \over < \int \sqrt{g} > }
\mathrel{\mathop\sim_{ V \rightarrow \infty }}
{ 1 \over V^{\delta / ( 1 + \delta ) } } ,
\eeq
with $\delta \approx 0.63$.
It is reasonable to assume that the exponent $\omega$ is the same in the
two ensembles, in which case one gets $\omega \approx 2.60 $.
But this result then implies that the volume fluctuations cannot drive
a continuous phase transitions. If this were the case, then the specific heat
exponent $\alpha \equiv 2 - 4 \nu = 1 + \omega $ would have to be
$\alpha < 1 $ or $\nu > 1/d = 1/4$, otherwise the transition is
expected to be first order \cite{nn},
in which case one would not be able to define a lattice continuum limit.
Indeed a direct determination of the volume fluctuations shows that
they are always finite, and in particular do not diverge at the
critical point at $k_c$, indicating that the mass associated with
the volume fluctuations (the conformal mode) is of the order of the
ultraviolet cutoff \cite{tala,ph1}.

Let us look for completeness at the analogous result for the curvature
distribution. Again the exponents appearing in this case can be
related to the curvature critical exponent $\delta$.
Let us assume, as seems to be the case, that close to the critical
point $k_c$ one has
\beq
\cR (k) \equiv { < \int \sqrt{g} R > \over < \int \sqrt{g} > }
\sim + { 1 \over V } {\partial \over \partial k } \ln Z
\mathrel{\mathop\sim_{ k \rightarrow k_c }} - A_{\cR} ( k_c - k )^\delta  .
\eeq
(see Eq. (\ref{eq:rsing})).
Then for the curvature fluctuation one expects close to $k_c$
\beq
\chi_{\cR} \sim { 1 \over V } {\partial^2 \over \partial k^2 } \ln Z
\mathrel{\mathop\sim_{ k \rightarrow k_c }}
{ \delta A_{\cR} \over ( k_c - k )^{1 - \delta} } .
\eeq
Here we are interested in the singular part of the free energy.
Close to the singularity the partition function is then given by
\beq
Z_{sing} ( k ) \sim \exp \left \{ - V \int^k d k' \,
A_{\cR} ( k_c - k' )^\delta + {\rm reg.}  \right \} .
\eeq
Now let us introduce the quantity $N(R)$ defined by
\beq
N (R) = { 1 \over 2 \pi i } \int_{- i \infty }^{ + i \infty }
d k \, Z ( k ) \, e^{ k R } ,
\eeq
with $ R = - V \cR $ ($R$ is therefore a positive quantity, related to
the magnitude of the curvature,  in the smooth phase where $\cR < 0 $).
In the saddle point approximation the density
of states is given by
\beq
N ( R ) \sim \exp \left \{ k_c R -
{\textstyle {\delta \over 1 + \delta} \displaystyle } \, R \,
[ R / ( V A_\cR ) ]^{ 1 / \delta }  \right \} .
\eeq
We find therefore that the full probability distribution for $R$
has an algebraic singularity close to $R=0$ of the type
\beq
\ln P(R) \equiv - k R + \ln N ( R ) \sim
( k_c - k) R - {\textstyle {\delta \over 1 + \delta} \displaystyle }
\, R \, [ R / ( V A_\cR ) \, ]^{ 1 / \delta } .
\eeq
Again there will also be a regular part, which we have omitted here.
One can verify that the stationary point of the distribution $P(R)$
gives indeed the singular behavior of Eq. (\ref{eq:rsing}).

\vskip 10pt
\newsection{Conclusions}

In the previous sections we have presented some first results
regarding the effects of scalar matter on quantized gravity, in the
context of a quantum gravity model based on Regge's simplicial
formulation. It was found that the feedback of the scalar fields on
the geometry is quite
small on purely gravitational quantities such as the average
curvature, in agreement with some of the perturbative predictions
in the continuum, which also seem to suggest that the scalar vacuum
polarization effects should be rather small.
The qualitative features of the phase diagram for gravity, and in
particular the appearance of a smooth and a rough phase, seem
unchanged, at least as long as one does not have too many matter fields.
It appears therefore that the approximation in which matter internal
loops are neglected (quenched approximation) could be considered a
reasonable one, and that quantities such as the critical exponents
should not be too far off in this case.
To the extent that the coupling between the scalar and metric
degrees of freedom is weak close to the critical point, we have argued
that gravity is indeed weak, and have presented a procedure by
which the effective low energy Newton's constant can be estimated
independently of the renormalized cosmological constant, which is
determined from the scaling behavior of the average curvature close to the
critical point.
Our results suggest that in this model the effective
gravitational coupling close to the ultraviolet fixed point grows
with distance, and is expected to depend in a
non-trivial way on the overall linear size of the system.
For the gravitational coupling we have found an infrared growth
away from the fixed point of the
type $ G ( \mu ) \sim \mu^{- 1 / \nu }$, while for the cosmological
constant we have found a decrease in the infrared,
$ \Lambda (\mu ) \sim \mu^{ 4 - 1 / \nu }$, with an exponent
$\nu$ given approximately by $\nu \approx 0.41$, and only weakly
dependent on the matter content.

Finally let us add that our results bear some similarity with the
results obtained recently from the dynamical
triangulation model in four dimensions \cite{dts}, where the scalar
field also seems to give a rather small contribution. On the other
hand the matter contribution does not seem to improve on the fact
that in these models, which only allow discrete local curvatures,
the average curvature does not show the correct scaling behavior
close to the critical point,
which is a necessary condition for defining a lattice continuum limit
(in these models at the critical point the curvature diverges in
physical units).
Clearly more work is needed in both models to further clarify these
issues.

\vspace{12pt}

{\bf Acknowledgements}

The numerical computations were performed at the NSF-sponsored SDSC, NCSA
and PSC Supercomputer Centers under a {\sl Grand Challenge} allocation grant.
The parallel MIMD version of the quantum gravity program was written
and optimized for the CM5-512 with Yasunari Tosa of TMC, and his
invaluable help is here gratefully acknowledged.

\newpage

\noindent {\large \bf Figure Captions}

\vspace{12pt}

\begin{itemize}

\item[Fig.\ 1]
Average curvature $\cR $ as a function of the mass of the scalar field
$m$, for different values of $k = 1 / 8 \pi G $. From top to bottom
$k=$ 0.0, 0.05, 0.1, 0.15, 0.2. The values for pure gravity ($z=0$) are
included for comparison, and drawn also as lines of constant $\cR$.
The values for $m=1.0$ ($z=0.5$) and $m=0.2$ ($z=0.962$) are from
a relatively small lattice with $L=4$ and are therefore for reference only,
while the values for $m=0.5$ ($z=0.80$) are
averages from the $L=8$ and $L=16$ lattices, with much smaller uncertainties.
The slight but clear decrease in the magnitude of the curvature in the
presence of the scalar field should be noted.

\item[Fig.\ 2]
Comparison of the average curvature $\cR$ as a function of $k$ in
the presence ($\diamond$) and absence ($\Box$) of the scalar field, with
mass $m=0.5$. The results for pure gravity are from ref. ~\cite{ph1} on
an $L=16$ lattice. The line corresponds to a fit of the pure gravity
results to an algebraic singularity, as discussed in the text.

\item[Fig.\ 3]
Minus the average curvature $\cR$ raised to the power
$1 / \delta = 1/0.63 $.
Parameters and data are the same as in Fig. 2. The straight line is
a fit to the pure gravity results. The linearity is now quite striking.

\item[Fig.\ 4]
Difference $\Delta \cR (k) $ between the average curvature in the presence
and absence of one scalar field, again for $m=0.5$ and $L=8,16$.
The difference is small and positive. The curve represents a
behavior close to the critical point of the type
$ \Delta \cR (k) \sim A \, ( k_c - k )^\delta $, with
$\delta \approx 0.63 $ and $k_c \approx 0.244 $ (the values for
pure gravity).

\item[Fig.\ 5]
The scalar field average $<\phi^2>$ as a function of $m$, and
for different values of the bare gravitational coupling $k$
($k=0.0,0.05,0.10,0.15,0.20$).
The data for $m=1.0$ and $m=0.2$ is from a lattice with $L=4$,
while data for $m=0.5$ from lattices with $L=8$ and 16.
The line is a fit assuming the free-field dependence on the mass $m$.

\item[Fig.\ 6]
Same as in Fig. 5, but for the scalar field average $<\phi^4>$.

\item[Fig.\ 7]
Same as in Fig. 5, but for the scalar field fluctuation $\chi_\phi $.

\item[Fig.\ 8]
A possible schematic phase diagram for gravity coupled to $n_f$ scalar
fields. The presence of the scalar fields shifts the critical point
$k_c = 1 / 8 \pi G_c $ towards smaller values as the number of scalar
flavors is increased, until the smooth phase disappears entirely for some
large number of flavors.

\end{itemize}

\end{document}